# Towards a reliable prediction of the infrared spectra of cosmic fullerenes and their derivatives in the JWST era

Jianzhi Xu[1], Aigen Li[2], Xiaohu Li[3], Gao-Lei Hou[1]

1. MOE Key Laboratory for Non-Equilibrium Synthesis and Modulation of Condensed Matter, School of Physics, Xi´an Jiaotong University, Xi´an, 710049 Shaanxi, China; gaolei.hou@xjtu.edu.cn

2. Department of Physics and Astronomy, University of Missouri, Columbia, MO 65211, USA; lia@missouri.edu

3. Xinjiang Astronomical Observatory, Chinese Academy of Sciences, Urumqi 830011 Xinjiang, China

**ABSTRACT**

Fullerenes, including $C_{60}$, $C_{70}$, and $C_{60}^+$, are widespread in space through their characteristic infrared vibrational features ($C_{60}^+$ also reveals its presence in the interstellar medium through its electronic transitions) and offer great insights into the carbon chemistry and stellar evolution. The potential existence of fullerene-related species in space has long been speculated and recently put forward by a set of laboratory experiments of $C_{60}^+$, $C_{60}H^+$, $C_{60}O^+$, $C_{60}OH^+$, $C_{70}H^+$, and $[C_{60}\text{-Metal}]^+$ complexes. The advent of the James Webb Space Telescope (JWST) provides a unique opportunity to search for these fullerene-related species in space. To facilitate JWST search, analysis, and interpretation, an accurate knowledge of their vibrational properties is essential. Here, we compile a VibFullerene database and conduct a systematic theoretical study on those species. We derive a set of range-specific scaling factors for vibrational frequencies, to account for the deficiency of density functional theory calculations in predicting the accurate frequencies. Scaling factors with low root-mean-square and median errors for the frequencies are obtained, and their performance is evaluated, from which the best-performing methods are recommended for calculating the infrared spectra of fullerene derivatives which balance the accuracy and computational cost. Finally, the recommended vibrational frequencies and intensities of fullerene derivatives are presented for future JWST detection.

## 1. INTRODUCTION

The discovery of fullerene in 1985, motivated by Sir Harold Kroto's curiosity about the formation mechanism of carbon-chain molecules in circumstellar envelopes, was a serendipitous event (Avery *et al.* 1976, Kroto *et al.* 1978, Broten *et al.* 1978, Kroto *et al.* 1985). Later macroscopic synthesis of $C_{60}$ enabled its spectroscopic characterization in both solid and vapor phases, providing laboratory reference to search for $C_{60}$ potentially in space, as speculated by Kroto based on the high stability of $C_{60}$ in harsh radiation conditions (Kroto et al. 1985, Kroto 1988). In 2010, Cami and co-workers made the first identification of circumstellar $C_{60}$ through its infrared-active features at 7.0, 8.5, 17.4, and 18.9 µm in a young planetary nebula Tc-1 (Cami *et al.* 2010). Over the past decade, $C_{60}$ has been detected in various astrophysical environments, including evolved stars, reflection nebulae, photodissociation regions, and the diffuse interstellar medium (ISM) (García-Hernández *et al.* 2010, García-Hernández *et al.* 2011b, García-Hernández *et al.* 2011a, García-Hernández *et al.* 2012, Gielen *et al.* 2011, Zhang & Kwok 2011, Bernard-Salas *et al.* 2012, Castellanos *et al.* 2014, Sellgren *et al.* 2010, Peeters *et al.* 2012, Berné *et al.* 2017), suggesting that it is widely and abundantly spread in space (Woods 2020).

Fullerenes may also reveal their presence in space through their characteristic electronic transitions. Indeed, in 2015, Maier and co-workers measured the first high-resolution, gas-phase absorption spectrum of $C_{60}^+$ utilizing their newly developed cryogenic ion trap via helium-tagging technique (Campbell *et al.* 2015), and identified two strong features at 9577 and 9632 Å and three weak features at 9348, 9365, and 9428 Å (Campbell *et al.* 2016b, Campbell et al. 2015, Campbell *et al.* 2016a, Campbell & Maier 2018), coincident with the mysterious diffuse interstellar bands at 9365, 9428, 9577, and 9632 DIBs as confirmed in 2019 by Hubble Space Telescope (Cordiner *et al.* 2019, Walker *et al.* 2016).

To date, $C_{60}$, $C_{60}^+$, and $C_{70}$ are the largest molecules identified in space. $C_{60}$ only has four active infrared features due to its $I_h$ symmetry. Distorting the carbon cage by ionization or adsorption of foreign atoms or groups could reduce the symmetry and introduce more active infrared features. In fact, fullerenes undergo complicated chemical processing during stellar evolution, making the formation of hydrogenated fullerenes, fullerene-metal and fullerene-hydrocarbon complexes possible (Kroto & Jura 1992, Cami 2014, Zhang *et al.* 2020, Hou *et al.* 2023, Barzaga *et al.* 2022, Omont 2016, Zhang *et al.* 2022). Due to its high spectral resolution and unprecedented sensitivity, the advent of the James Webb Space Telescope (JWST) offers a valuable opportunity to detect fullerene derivatives through their vibrational bands. This will allow us to quantitatively determine (or place an upper limit on) the presence and abundances of specific derivatives of fullerene in space. To this end, an accurate knowledge of the frequencies and intensities of the vibrational bands of fullerene derivatives is required.

Also, Gerlich et al. (2018) recently reported the gas-phase infrared spectrum of $C_{60}^+$ using He-tagging technique in the spectral range of 1100–1600 cm$^{-1}$, but without detailed explanation on the origin of the measured features. Oomens and co-workers measured the infrared spectra of gas-phase $C_{60}H^+$, $C_{60}O^+$, $C_{60}OH^+$, and $C_{70}H^+$, using the infrared multiple photon dissociation (IRMPD) spectroscopy (Palotás *et al.* 2019, Palotás *et al.* 2022, Palotás *et al.* 2021) and explored their possible relevance to the so-called "unidentified infrared emission" (UIE) bands which are ubiquitously seen at 3.3, 6.2, 7.7, 8.6, and 11.3 µm in a wide range of astrophysical regions. These elegant but challenging laboratory experiments are essential to fully understand the astronomical observations. Theoretical simulations, in particular those based on density functional theory (DFT), have proved to be a superior complementary approach, on one hand to reveal the feature origin by comparing with experiment, and on the other hand to predict the infrared spectra of experimentally unstudied species (Scott & Radom 1996, Neese 2009, Sadjadi *et al.*

2022, Kerkeni *et al.* 2022, Candian *et al.* 2019). The prediction can either base on harmonic approximation, or use the more accurate but also more expensive anharmonic method. To achieve a reliable prediction using the much cheaper harmonic calculations, scaling factors have been developed for correcting the computed frequencies (Pople *et al.* 1993, Scott & Radom 1996, Zapata Trujillo & McKemmish 2023).

In this work, we systematically investigate the infrared spectral properties of several experimentally measured fullerene-related species, including $C_{60}$, $C_{70}$, $C_{60}^+$, $C_{60}O^+$, $C_{60}OH^+$, $C_{60}H^+$, and $C_{70}H^+$ (hereafter we call them the VibFullerene database), and compare them with those computed using several popular DFT methods. We provide a set of scaling factors for a variety of DFT methods that lead theoretical calculations to better match the experiments, which might be employed in the JWST era to compare with observational data to explore their potential presence in space. By making use of appropriate scaling factors, we aim to simulate the infrared spectra of fullerene derivatives with high reliability and low cost. Readers who are interested only in the vibrational frequencies and intensities for future JWST detection may wish to proceed directly to Section 5.

## 2. DATABASE COMPILATION

A benchmark database containing gas-phase infrared spectra is essential for obtaining the scaling factors and thus the simulation of other related species. Pople et al. previously built a database of scaling factors for the calculations at HF, MP2, QCISD, and DFT levels (Pople et al. 1993), Truhlar et al. collected a database for scaling the calculated vibrational frequencies (Alecu *et al.* 2010), and recently McKemmish and co-workers compiled an updated database, VIBFREQ1295 for the scaling factors of vibrational frequency calculations (Zapata Trujillo & McKemmish 2022). Those databases mainly contain small molecules of fewer than 15 atoms. In this work, we compile in Table 1 a VibFullerene database consisting of 7 fullerene-related species, i.e., $C_{60}$, $C_{70}$, $C_{60}^+$, $C_{60}O^+$, $C_{60}OH^+$, $C_{60}H^+$, and $C_{70}H^+$, totally 92 vibrational frequencies. The database contains information on the chemical formula, experimental method, vibrational frequencies, and the number of frequencies considered. We use the notation 92f/7mol to describe the VibFullerene database, where f and mol represent the total number of frequencies and molecules considered. We note here that $C_{60}$ and $C_{70}$ are from vapor phase FTIR measurements (JohnáDennis *et al.* 1991, Schettino *et al.* 2002), while others are from gas-phase infrared dissociation experiments.

**Table 1.** Summary of fullerene-related species considered in VibFullerene (92f/7mol) with chemical formulas, experimental methods, vibrational frequencies, number of frequencies, and references.

| Formula | Method | Frequencies | Number | Ref. |
|---|---|---|---|---|
| $C_{60}$ | FTIR | 527, 577, 1183, 1429 | 4 | (JohnáDennis et al. 1991) |
| $C_{70}$ | FTIR | 533, 565, 575, 642, 669, 795, 1087, 1134, 1163, 1205, 1241, 1295, 1325, 1416, 1428, 1462, 1490, 1563 | 18 | (Schettino et al. 2002) |
| $C_{60}^+$ | IR(M)PD | 1173, 1178, 1191, 1216, 1222, 1307, 1327, 1402, 1544, 1560 | 10 | (Gerlich et al. 2018) |
| $C_{60}H^+$ | IRMPD | 524, 566, 756, 953, 1037, 1094, 1167, 1199, 1229, 1301, 1370, 1415, 1467, 1523, 1545 | 15 | (Palotás et al. 2019) |
| $C_{60}O^+$ | IRMPD | 519, 952, 1067, 1088, 1181, 1240, 1285, 1333, 1390, 1516 | 10 | (Palotás et al. 2022) |
| $C_{60}OH^+$ | IRMPD | 526, 573, 705, 760, 786, 958, 1029, 1088, 1173, 1195, 1232, 1296, 1328, 1411, 1468, 1526, 1540 | 17 | (Palotás et al. 2022) |
| $C_{70}H^+$ | IRMPD | 528, 570, 636, 665, 685, 723, 795, 821, 893, 933, 1065, 1160, 1208, 1304, 1368, 1434, 1498, 1543 | 18 | (Palotás et al. 2021) |

The VibFullerene database can be used to investigate the distortion effect on the vibrational frequencies of fullerenes, and to validate theoretical models for calculating vibrational frequencies. It should be pointed out that due to the experimental challenge in measuring the gas-phase infrared spectra of large molecules like fullerenes and their derivatives, VibFullerene database currently only contains 7 molecules. We expect to expand the database to include more species in the future (Vanbuel *et al.* 2020, Hou *et al.* 2021, Li *et al.* 2022, Hou et al. 2023, German *et al.* 2022).

**3. METHODOLOGY**

All geometries in the VibFullerene database are fully optimized without symmetry constriction, using 8 different DFT functionals, including BP86 (Becke 1988, Perdew 1986), BPW91 (Becke 1988, Perdew & Wang 1992), TPSSh (Tao *et al.* 2003), PBE0 (Adamo & Barone 1999), B3LYP (Stephens *et al.* 1994), M06 (Zhao & Truhlar 2008), M06-2X (Zhao & Truhlar 2008), and ωB97XD (Chai & Head-Gordon 2008a, Chai & Head-Gordon 2008b). Since vibrational frequencies are less affected by basis sets than by the specific functional (Pagliai *et al.* 2014), the geometric parameters of the species in VibFullerene were compared using B3LYP functional in combination with several different basis sets, including 3-21G (Binkley *et al.* 1980), 4-31G (Ditchfield *et al.* 1971), 6-31G** (Ditchfield et al. 1971, Hehre *et al.* 1972, Hariharan & Pople 1973), 6-311G** (Krishnan *et al.* 1980), def2-SVP (Weigend & Ahlrichs 2005), and N07D (Barone *et al.* 2008), to choose the best basis set in the following frequency calculations considering both accuracy and efficiency. Note that 4-31G and N07D have been widely employed to investigate the infrared features of cosmic polycyclic aromatic hydrocarbons (PAHs). Frequency analysis was carried out to assure the optimized geometries to be real minima and to simulate the theoretical infrared spectra. Anharmonic spectrum was computed by generalized second order vibrational perturbation theory (GVPT2) method (Barone 2005, Barone *et al.* 2014). All calculations were conducted with Gaussian 09 program package (Frisch *et al.* 2013).

Scaling factors (λ) were determined by a least-square fitting procedure via minimizing the root mean square error (RMSE) (Pople et al. 1993, Scott & Radom 1996)

$$\lambda = \sum_{i}^{N} \omega_{i,\ theo} \tilde{\nu}_{i,\ exp} / \sum_{i}^{N} (\omega_{i,\ theo})^2 \quad (1)$$

where $\omega_{i,\ theo}$ and $\tilde{\nu}_{i,\ exp}$ are $i$th theoretical and experimental frequencies, respectively, and N is the sum of all the considered vibrational modes in Table 1.

To evaluate the accuracy of the fitted scaling factors, we calculated the median errors as a more robust evaluation compared to RMSE, which can be significantly influenced by outliers (Zapata Trujillo & McKemmish 2022, Zapata Trujillo & McKemmish 2023). Specifically, median errors refer to the median of the differences between the $\tilde{\nu}_{i,\ exp}$ and $\lambda\omega_{i,\ theo}$.

It is found that theoretical frequencies are generally less overestimated in the low-frequency range than in the high-frequency region compared to the experimental fundamentals. Hence, a frequency-range-specific scaling approach (Zapata Trujillo & McKemmish 2023) is recommended to improve the accuracy of calculations. This method involves identifying an optimal segmentation point to separate the low-frequency and high-frequency ranges. The segmentation point with the least median error to separate the spectra into low- and high-frequency regions is searched for.

To quantitatively assess the agreement between the calculated and experimental infrared spectra, we employed the recently proposed cosine similarity score as an objective measure (Fu & Hopkins 2018, Kempkes *et al.* 2019, Müller *et al.* 2020). The cosine of the angle θ between two *n*-dimensional vectors is calculated using their normalized Euclidean dot product according to

$$\text{Similarity} = \cos(\theta) = \frac{\mathbf{A}\cdot\mathbf{B}}{||\mathbf{A}||\,||\mathbf{B}||} = \frac{\sum_{i=1}^{n} A_i B_i}{\sqrt{\sum_{i=1}^{n} A_i^2}\cdot\sqrt{\sum_{i=1}^{n} B_i^2}} \quad (2)$$

where **A** and **B** are two *n*-dimensional vectors with $A_i$ and $B_i$ as their $i^{th}$ elements. This method assesses the degree to which the two vectors, representing the experimental and theoretical spectra in this case, are parallel. A cosine value closer to 1 indicates higher similarity. The intensity values in the computed spectrum are evaluated at the exact wavenumbers of the experimental spectrum, so that the two spectra have a common x-axis. To consider possible scaling for the calculated harmonic frequencies, the calculated spectra can be scaled first before calculating the Cosine similarity scores.

Kempkes et al. proposed a slightly modified version of the above formula to make the cosine similarity scores more sensitive to the frequency overlap between bands in ***A*** and ***B***, and less to the deviations in their peak intensities (Kempkes et al. 2019). Both the experimental and calculated spectra are scaled to a maximum intensity of 1 and then the logarithm of these scaled values is taken as

$$\boldsymbol{A}_i^{rev} = \log\left(\frac{\boldsymbol{A}_i}{\boldsymbol{A}_{max}} + c\right) \quad (3)$$

where *c* is a constant that is identical for vectors ***A*** and ***B***. The value of *c* is a compromise between being sensitive to low-intensity bands in the spectrum on the one hand and avoiding experimental noise affecting the similarity on the other hand. We followed this approach in the current work and used a *c* value of 0.71 by testing a small set of experimental and computational spectra to give the best results.

## 4. RESULTS

### 4.1 Geometric parameters

Table 2 compares the characteristic bond lengths of $C_{60}$, $C_{70}$, $C_{60}H^+$, $C_{60}O^+$, and $C_{70}H^+$ computed with different methods with those measured experimentally or computed theoretically (Hedberg *et al.* 1991, Hedberg *et al.* 1997, Palotás et al. 2019, Palotás et al. 2022, Palotás et al. 2021). It is shown that all the six basis sets are reasonable in reproducing the geometries of $C_{60}$ and $C_{70}$, except that the experimental bond length of bond 8 of $C_{70}$ is much longer (the notation of $C_{70}$ bonds can be found in Figure A1 of the Appendix). That is probably due to the elongation of equatorial bonds at high temperatures as employed in the experimental measurement.

It is noted that the polarization function is crucial for obtaining reliable structural results (Xu & Hou 2022), and 3-21G and 4-31G basis sets can lead to errors up to 0.03 Å, in particular for $C_{60}O^+$ in the calculations. Among the six basis sets, def2-SVP induces an error of approximately 1% when H atom is involved, as in the case of $C_{60}H^+$ and $C_{70}H^+$. The basis sets of 6-31G**, 6-311G**, and N07D, all yield satisfactory results. However, considering the computational cost of frequency analysis (Table 2), the 6-31G** basis set is recommended. This basis set was previously found to produce only minor errors in predicting the Raman and infrared modes of $C_{60}$ (Pagliai et al. 2014).

**Table 2.** Comparison of characteristic bond length (Å) and computation cost (in seconds, s) of vibrational analysis of fullerene-related species in VibFullerene using B3LYP functional combined with different basis sets.

| Species | Bond[a] | 3-21G | 4-31G | 6-31G** | 6-311G** | def2-SVP | N07D | Reference values |
|---|---|---|---|---|---|---|---|---|
| $C_{60}$ | C–C | 1.460 | 1.457 | 1.453 | 1.451 | 1.455 | 1.454 | 1.458(6)[b] |
| | C=C | 1.390 | 1.393 | 1.396 | 1.392 | 1.398 | 1.397 | 1.401(10)[b] |
| $C_{70}$ | 1 | 1.459 | 1.455 | 1.452 | 1.450 | 1.453 | 1.453 | 1.461(8)[c] |
| | 2 | 1.392 | 1.394 | 1.397 | 1.394 | 1.400 | 1.398 | 1.388(17)[c] |
| | 3 | 1.455 | 1.451 | 1.448 | 1.446 | 1.450 | 1.449 | 1.453(11)[c] |
| | 4 | 1.384 | 1.386 | 1.389 | 1.385 | 1.391 | 1.390 | 1.386(25)[c] |
| | 5 | 1.456 | 1.452 | 1.450 | 1.447 | 1.451 | 1.450 | 1.468(11)[c] |
| | 6 | 1.437 | 1.436 | 1.434 | 1.431 | 1.436 | 1.435 | 1.425(14)[c] |
| | 7 | 1.419 | 1.420 | 1.421 | 1.418 | 1.423 | 1.422 | 1.405(13)[c] |
| | 8 | 1.474 | 1.473 | 1.471 | 1.469 | 1.472 | 1.472 | 1.538(19)[c] |
| $C_{60}H^+$ | C–H | 1.106 | 1.105 | 1.105 | 1.103 | 1.112 | 1.106 | 1.103[d] |
| $C_{60}O^+$ | C–O | 1.401 | 1.400 | 1.374 | 1.372 | 1.366 | 1.372 | 1.37[e] |
| $C_{70}H^+$ | C–H | 1.106 | 1.105 | 1.105 | 1.104 | 1.112 | 1.107 | 1.104[f] |
| Time (s)[g] | | 740 | 915 | 3189 | 9427 | 6692 | 58999 | |

[a]Notation of bond parameters of $C_{70}$ is provided in Figure A1 of the Appendix. [b,c]Electron diffraction data from refs.(Hedberg et al. 1991, Hedberg et al. 1997) [d,e,f]Computational results from refs.(Palotás et al. 2019, Palotás et al. 2021, Palotás et al. 2022) [g]Time is evaluated by frequency analysis of $C_{60}$ on a 52-core SE24HIJTP workstation.

### 4.2 Scaling factors

Theoretical infrared spectra based on harmonic approximation neglect anharmonic effects and often deviate from that experimental measurements for most hybrid functionals, and applying a uniform scaling factor in most cases could be able to correct the effect to a satisfactory extent (Pople et al. 1993). In Figure 1, the distribution of differences between the calculated harmonic frequencies and experimental fundamentals are presented. For both BP86 and BPW91, the errors seem constant throughout the 400–1600 $cm^{-1}$ spectral range, while for the other six hybrid functionals, the differences increase almost linearly with increasing wavenumber, calling for careful scaling of the calculated frequencies.

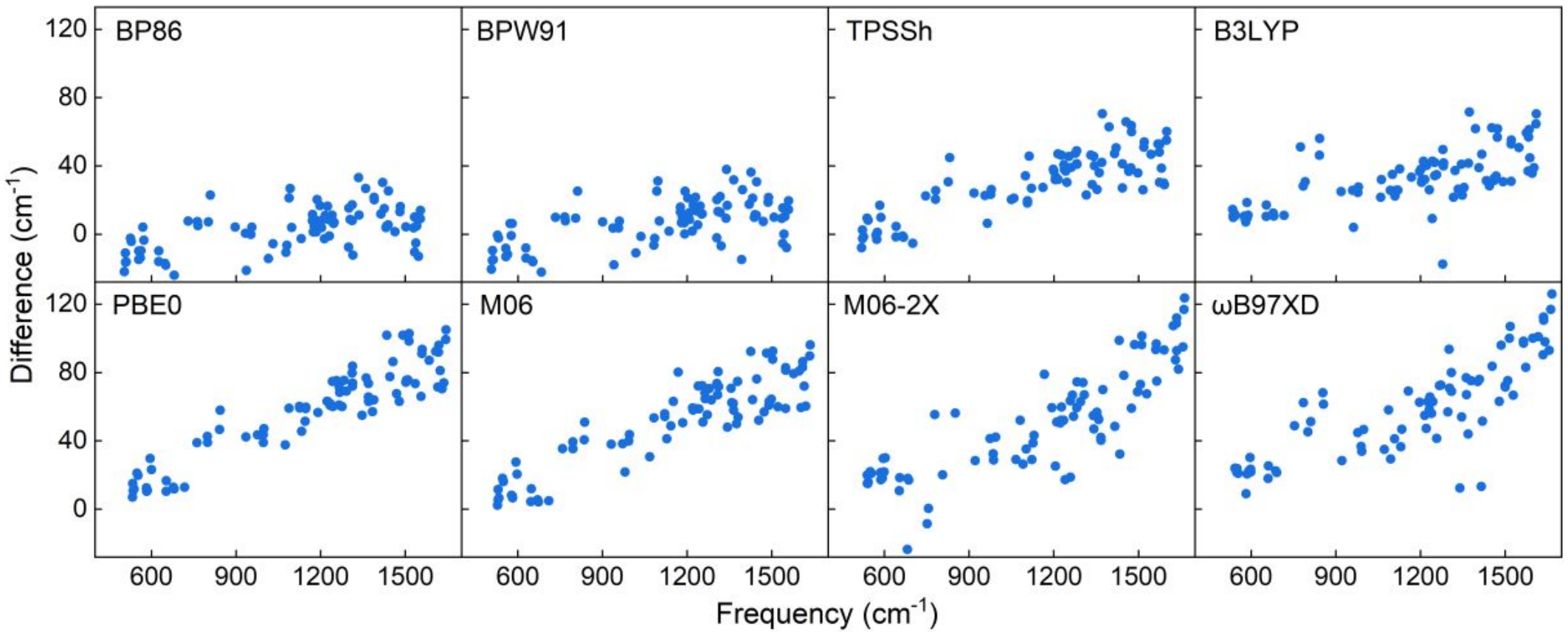

**Figure 1.** Distribution of the differences between the calculated harmonic and experimental frequencies in the full investigated spectral range. All calculations are performed with 6-31G** basis set.

To obtain better agreement between the simulation and experiment, frequency-range-specific scaling factors would be desired as seen from the variations of the frequency differences in Figure 1. We therefore fit sequentially in steps of 50 $cm^{-1}$ over the entire spectral range to find the best segmentation point, and the results are summarized in Table A1 of the Appendix. It is shown that, compared with universal scaling factors, the median errors and root mean square errors (RMSE) are significantly reduced when the frequency-range-specific strategy is employed. We choose the segmentation point with the least median error to separate the spectra into low- and high-frequency regions. Note that some segmentation points with the least median error do not have the least RMSE values as RMSE is sensitive to outliers as explained in the Methodology section. The scaling factors are thus fitted separately for low- and high-frequency regions (Figure 2), and the best-fitted scaling factors are summarized in Table 3. All $R^2$ values are larger than 0.997, suggesting satisfactory fittings. Note that some significant errors, such as that for the 795 $cm^{-1}$ band of $C_{70}$, are not included and will be discussed in detail next. We note that the scaling factors obtained here for the high-frequency range are consistent with that previously obtained for small molecules (Scott & Radom 1996, Merrick *et al.* 2007, Alecu et al. 2010, Russell & Johnson 2022). It is also noted that the functionals with higher Hartree-Fock (HF) components tend to overestimate the calculated harmonic frequencies more, necessitating smaller scaling factors. For instance, the scaling factors for both BP86 and BPW91 in either low- or high-frequency range are close to unity, while for M06-2X with 54% HF component, scaling factors of 0.9631 and 0.9485 are required for the low- and high-frequency ranges, respectively.

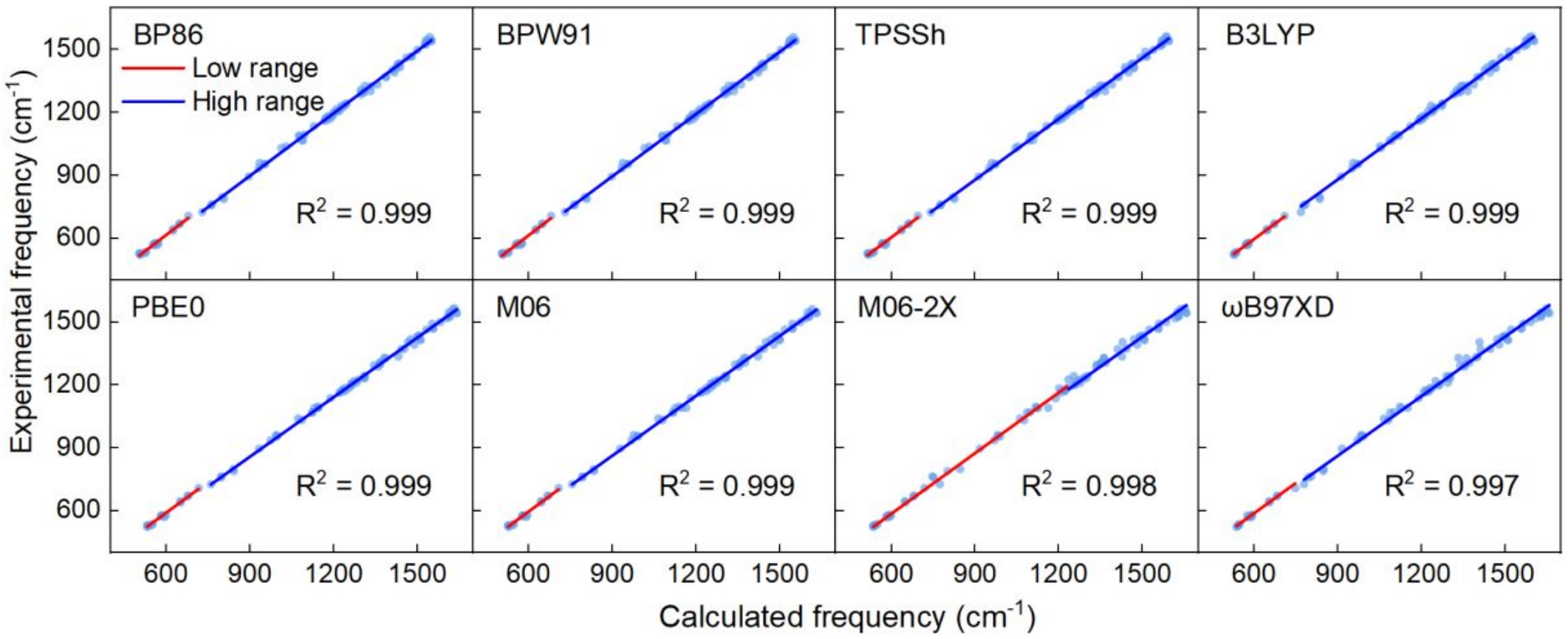

**Figure 2**. Scaling factors for the low- and high-frequency ranges of different functionals with 6-31G** basis set.

**Table 3**. Scaling factors obtained for different functionals with 6-31G** basis set.

| Functionals (HF%) | Scaling factors | | RMSE | Med. Error | Time (s) |
|---|---|---|---|---|---|
| BP86 (GGA) | 1.0212 for <700 $cm^{-1}$ | 0.9941 for >700 $cm^{-1}$ | 10.2 | 5.3 | 1351 |
| BPW91 (GGA) | 1.0165 for <700 $cm^{-1}$ | 0.9905 for >700 $cm^{-1}$ | 10.7 | 5.5 | 1354 |
| TPSSh (10%) | 0.9972 for <700 $cm^{-1}$ | 0.9701 for >700 $cm^{-1}$ | 10.1 | 6.3 | 3321 |
| B3LYP (20%) | 0.9807 for <750 $cm^{-1}$ | 0.9697 for >750 $cm^{-1}$ | 11.0 | 5.2 | 3189 |
| PBE0 (25%) | 0.9755 for <750 $cm^{-1}$ | 0.9472 for >750 $cm^{-1}$ | 9.6 | 6.3 | 3390 |
| M06 (27%) | 0.9833 for <750 $cm^{-1}$ | 0.9519 for >750 $cm^{-1}$ | 10.6 | 7.0 | 4184 |
| M06-2X (54%) | 0.9631 for <1250 $cm^{-1}$ | 0.9485 for >1250 $cm^{-1}$ | 16.4 | 8.3 | 3702 |
| ωB97XD | 0.9635 for <750 $cm^{-1}$ | 0.9475 for >750 $cm^{-1}$ | 16.8 | 8.5 | 4407 |

Table 3 suggests that overall functionals with low-HF components exhibit smaller RMSE and median error values, and Figure 3 presents the differences between the experimental fundamentals and the scaling-factor-corrected harmonic frequencies. It shows that either M06-2X or ωB97XD functionals may not be sufficiently robust, as a few frequencies exhibit deviations up to 40 $cm^{-1}$. A closer look at Figure 3 finds some functional-dependent large absolute differences for certain vibrational modes, for which the vibrational vectors are provided in Figure A2 of the Appendix. For example, BP86, BPW91, and M06 give the largest differences of 31, 26, and 25 $cm^{-1}$, respectively, for the coupling mode between tangential stretching of $C_{60}$ and bending of C–O–H in $C_{60}OH^+$ with an experimental frequency of 958 $cm^{-1}$. For TPSSh, the significant difference is 30 $cm^{-1}$ for the coupling mode between tangential stretching of $C_{60}$ and C–C–H bending in $C_{60}H^+$ (exptl. 1301 $cm^{-1}$); for B3LYP, it is 31 $cm^{-1}$ for the coupling mode between tangential stretching of $C_{60}$ and C–O–H bending in $C_{60}OH^+$ (exptl. 786 $cm^{-1}$); and for PBE0, it is 26 $cm^{-1}$ for the coupling mode between tangential stretching of $C_{60}$ and C–O–C symmetrical stretching in $C_{60}O^+$ (exptl. 1333 $cm^{-1}$). These results highlight the origin of the large errors is due to the deficiency of harmonic calculations in predicting the vibrational coupling modes in the analysis. Decomposing these coupling modes to assess the influence of heteroatoms on the cage vibrational modes would be a valuable practice, as we did previously (Hou et al. 2023), but is out of the scope of this work. Nonetheless, the errors due to the coupling modes generally fall in an acceptable range, emphasizing the potential of our approach to reliably predict the infrared spectra of fullerene-related species.

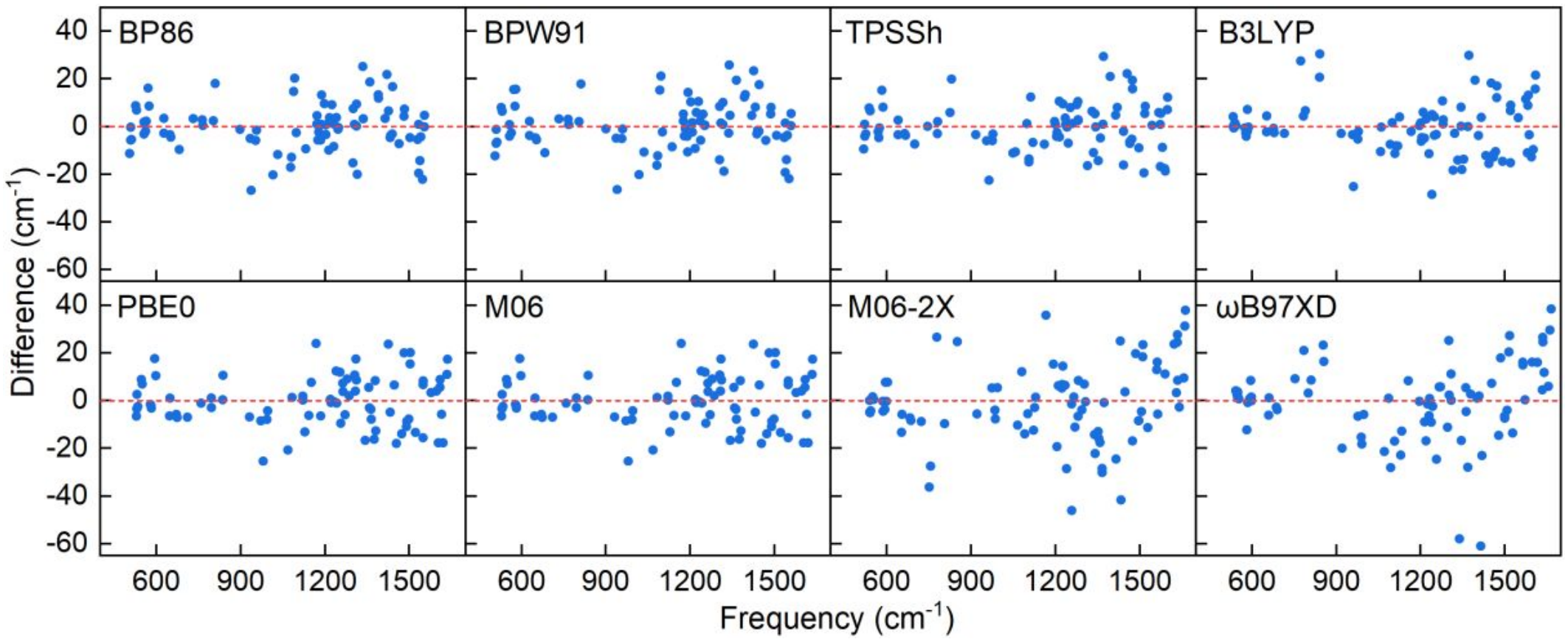

**Figure 3.** Distribution of frequency differences between the experimental fundamentals and the scaling-factor-corrected harmonic frequencies.

### 4.3 Verification of the performance of the fitted scaling factors

To test the performance of the fitted scaling factors, we generate a set of scaling-factor-corrected theoretical spectra and overlay them with the experimental spectra (Figures 4 and A3 of the Appendix). An explicit comparison of the performance of different functionals using the Cosine similarity score demonstrates the robustness of BP86 and BPW91 in predicting all fullerene-related species with high Cosine similarity (Fu & Hopkins 2018, Kempkes et al. 2019, Müller et al. 2020). For sake of clarity, only spectra calculated with BP86/6-31G** are presented here, and others can be seen in Figure A3 of the Appendix. Overall, good agreement is obtained between the corrected band frequencies and the experimental measurements except for several features of $C_{60}^{+}$ and $C_{70}$.

Figure 4 shows that the scaled harmonic spectrum of $C_{60}$ exhibits a good agreement with the FTIR measured spectrum regarding the vibrational frequencies, but the feature intensities exhibit some deviations. For $C_{60}H^{+}$ and $C_{60}O^{+}$, the theoretical spectra display stronger intensities around 520 $cm^{-1}$, which are identified as vibrational modes involving the radial breathing of the $C_{60}$ cage (vibrational vectors can be seen in Figure A2 of the Appendix). The theoretical spectrum of $C_{60}O^{+}$ also shows much stronger intensities than the IRMPD measurement in the spectral range of 1000–1500 $cm^{-1}$. The theoretical spectrum of $C_{60}OH^{+}$ shows large intensity deviations around 1400 $cm^{-1}$ which involves tangential stretching and around 1000 $cm^{-1}$ that mainly involves C–O stretching (Figure A2 of the Appendix). The differences in intensities between calculations and experiments could be attributed to the nonlinear effect due to multiple photon absorption, calling for more elegant infrared spectroscopy experiments in single-photon absorption scheme (Gerlich et al. 2018).

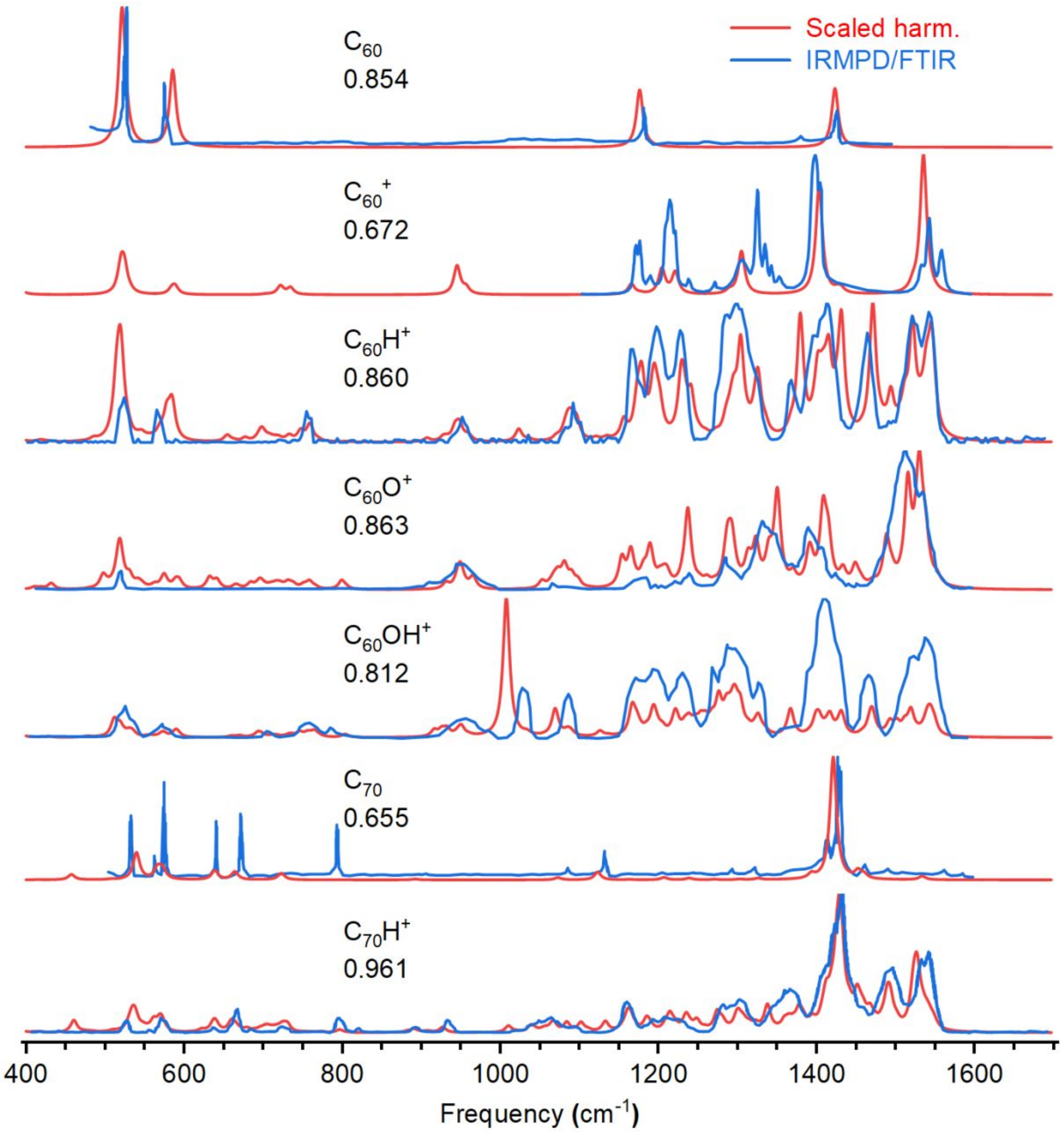

**Figure 4.** Scaled harmonic spectra at BP86/6-31G** and comparison with their experimental ones. The theoretical spectra are plotted using Lorentzian line shapes with full width at half maxima (FWHM) of 10 cm$^{-1}$. The Cosine similarity score for each functional is indicated.

For $C_{70}$, a large discrepancy is found for the equatorial mode between the experimental frequency of 795 cm$^{-1}$ and the theoretical value of 724 cm$^{-1}$ (Figure A2 of the Appendix), which has also been reported previously (Schettino et al. 2002, Stratmann *et al.* 1998, Pagliai et al. 2014, Palotás et al. 2021). Calculations including anharmonic effect fail to give a quantitative agreement on the frequency, presumably due to that the equatorial carbon atoms may be sensitive to temperature changes as higher temperatures during the experimental measurement may result in stronger vibrations of the equatorial belt compared to low temperatures (Hedberg et al. 1997).

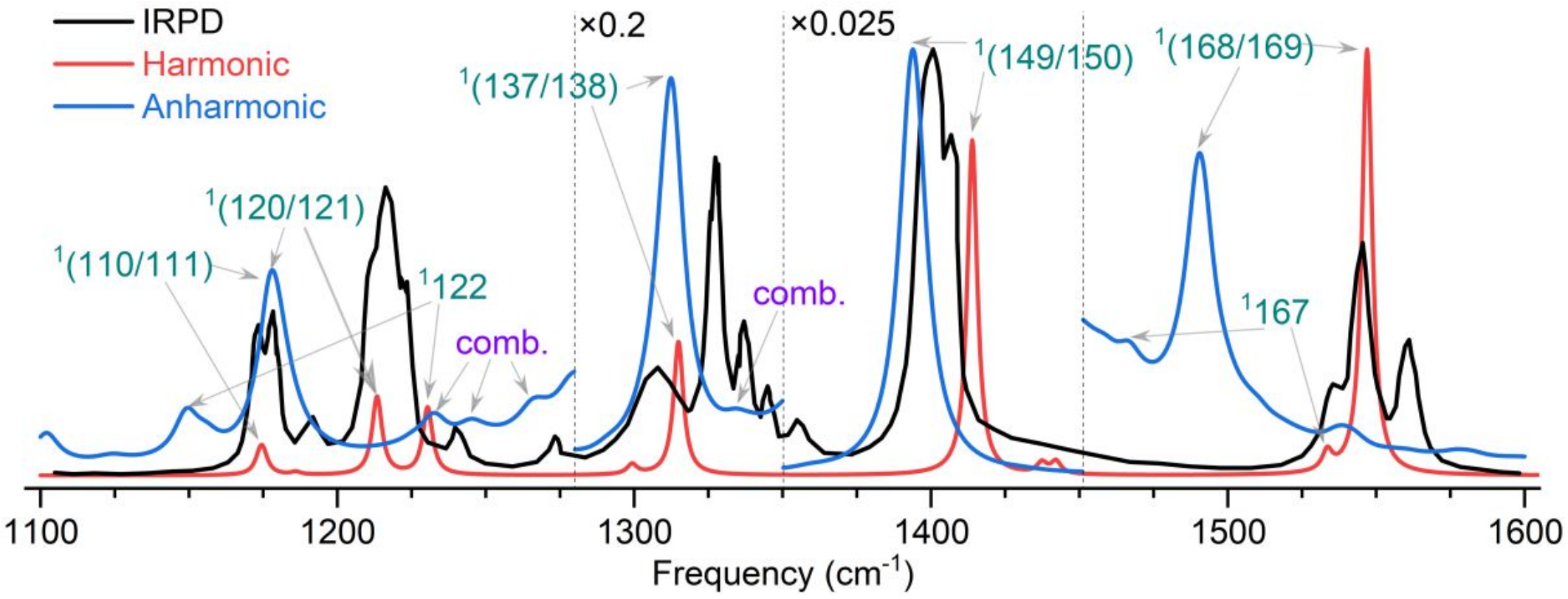

**Figure 5.** Simulated spectra of $C_{60}^+$ at BP86/6-31G** with harmonic approximation (red) and anharmonic effect (blue). The experimental infrared photodissociation (IRPD) spectrum is presented in black. The convolved anharmonic spectrum is scaled by a factor of 0.2 and 0.025 in the spectral range of 1280–1350 and 1350–1450 cm$^{-1}$, respectively. Tentative assignments of spectral features are provided with cyan and violet labels, corresponding to fundamental and combination bands, respectively.

The unsatisfactory comparison between the scaled harmonic spectrum of $C_{60}^+$ and experimental IRPD spectrum posts special attention as the experiment shows much richer features. Anharmonic effect could be one possible reason, and in Figure 5, we present the calculated anharmonic and harmonic spectra of $C_{60}^+$, both at the BP86/6-31G** level, in addition to the experimental spectrum. The comparison shows that the fundamentals from anharmonic calculation exhibit a redshift compared to that from the harmonic calculation. Due to the lack of high-quality potential energy surface for $C_{60}^+$, the fundamental frequencies from anharmonic calculations may not be superior to the scaled harmonic calculations, as shown by large differences in some modes. The experimental features in the spectral range of 1200–1400 cm$^{-1}$ may come from combination bands as suggested by the anharmonic calculations. However, the high cost of anharmonic calculation prohibits its application in general cases in predicting the infrared spectra of fullerene derivatives.

## 5. RECOMMENDATION FOR ASTROPHYSICAL APPLICATION

Table 2 demonstrates the suitability of 6-31G** basis set in predicting molecular geometries regarding both accuracy and computational cost, and detailed calculations with a range of DFT functionals suggest that functionals with low HF components, such as BP86, BPW91, TPSSh, B3LYP, PBE0, and M06, performs well in predicting the vibrational frequencies. In particular, the generalized gradient approximations (GGA), i.e., BP86 and BPW91, offer a good balance between the accuracy and computational efficiency, posting them interesting choices for simulating the infrared spectra of fullerene-related species. We note that the current work also validates the usefulness of the BPW91 functional in predicting the infrared spectra of [$C_{60}$-Metal]$^+$ (Metal = V and Fe) complexes (Hou et al. 2021, Hou et al. 2023). With the fitted scaling factors, excellent agreement is achieved between the simulated infrared spectra of $C_{60}V^+$ with laboratory measurements (Figure A3 of the Appendix).

**Table 4**. Recommended frequencies and intensities of fullerene derivatives in the spectral range of 400–1600 cm$^{-1}$ calculated at BP86/6-31G** level.

| Species | ν/cm$^{-1}$ | λ/μm | A/km·mol$^{-1}$ |
|---|---|---|---|
| $C_{60}$ | 522 | 19.2 | 22.1 |
| | 586 | 17.1 | 11.9 |
| | 1178 | 8.5 | 8.4 |
| | 1426 | 7.0 | 9.1 |
| $C_{70}$ | 458 | 21.8 | 6.9 |
| | 540 | 18.5 | 7.0 |
| | 567 | 17.6 | 7.3 |
| | 573 | 17.5 | 12.9 |
| | 639 | 15.6 | 6.0 |
| | 665 | 15.0 | 5.4 |
| | 724 | 13.8 | 3.9 |
| | 1074 | 9.3 | 1.5 |
| | 1124 | 8.9 | 9.9 |
| | 1209 | 8.3 | 1.9 |
| | 1241 | 8.1 | 1.2 |
| | 1305 | 7.7 | 0.4 |
| | 1328 | 7.5 | 1.9 |
| | 1396 | 7.2 | 3.0 |
| | 1424 | 7.0 | 79.2 |
| | 1463 | 6.8 | 2.6 |
| | 1536 | 6.5 | 2.1 |
| $C_{60}^+$ | 525 | 19.0 | 15.2 |
| | 589 | 17.0 | 5.3 |
| | 723 | 13.8 | 3.0 |
| | 736 | 13.6 | 2.4 |
| | 946 | 10.6 | 9.6 |
| | 1168 | 8.6 | 3.7 |
| | 1206 | 8.3 | 8.9 |
| | 1223 | 8.2 | 15.0 |
| | 1307 | 7.7 | 15.1 |
| | 1406 | 7.1 | 37.4 |
| | 1538 | 6.5 | 48.6 |
| $C_{60}H^+$ | 520 | 19.2 | 16.1 |
| | 547 | 18.3 | 1.3 |
| | 569 | 17.6 | 1.2 |
| | 578 | 17.3 | 6.2 |
| | 656 | 15.2 | 1.5 |
| | 748 | 13.4 | 2.5 |
| | 760 | 13.2 | 2.6 |
| | 949 | 10.5 | 3.2 |
| | 1025 | 9.8 | 3.2 |
| | 1097 | 9.1 | 4.7 |
| | 1157 | 8.6 | 4.6 |
| | 1175 | 8.5 | 8.1 |
| | 1180 | 8.5 | 11.2 |
| | 1195 | 8.4 | 9.9 |
| | 1231 | 8.1 | 14.6 |
| | 1306 | 7.7 | 23.1 |
| | 1328 | 7.5 | 14.7 |
| | 1381 | 7.2 | 30.8 |
| | 1403 | 7.1 | 13.2 |
| | 1418 | 7.1 | 10.9 |
| | 1433 | 7.0 | 24.8 |
| | 1474 | 6.8 | 17.8 |
| | 1523 | 6.6 | 27.3 |
| | 1548 | 6.5 | 17.9 |
| $C_{60}O^+$ | 432 | 23.1 | 2.3 |
| | 498 | 20.1 | 4.7 |
| | 518 | 19.3 | 8.6 |
| | 531 | 18.8 | 2.7 |
| | 575 | 17.4 | 3.9 |
| | 594 | 16.8 | 2.7 |
| | 633 | 15.8 | 4.1 |
| | 642 | 15.6 | 3.1 |
| | 684 | 14.6 | 2.1 |
| | 759 | 13.2 | 2.6 |
| | 800 | 12.5 | 3.2 |
| | 932 | 10.7 | 2.2 |
| | 951 | 10.5 | 7.3 |
| | 966 | 10.4 | 4.4 |
| | 1074 | 9.3 | 4.4 |
| | 1083 | 9.2 | 7.9 |
| | 1092 | 9.2 | 4.2 |
| | 1167 | 8.6 | 12.0 |
| | 1191 | 8.4 | 13.0 |
| | 1213 | 8.2 | 4.1 |
| | 1239 | 8.1 | 28.5 |
| | 1290 | 7.8 | 11.2 |
| | 1295 | 7.7 | 14.6 |
| | 1315 | 7.6 | 7.7 |
| | 1326 | 7.5 | 14.2 |
| | 1353 | 7.4 | 28.0 |
| | 1393 | 7.2 | 9.2 |
| | 1411 | 7.1 | 25.9 |
| | 1418 | 7.1 | 12.4 |
| | 1451 | 6.9 | 7.3 |
| | 1491 | 6.7 | 17.0 |
| | 1518 | 6.6 | 36.0 |
| | 1529 | 6.5 | 5.8 |
| | 1541 | 6.5 | 9.5 |
| $C_{60}OH^+$ | 519 | 19.3 | 10.0 |
| | 535 | 18.7 | 3.5 |
| | 575 | 17.4 | 2.2 |
| | 589 | 17.0 | 5.0 |
| | 670 | 14.9 | 1.7 |
| | 694 | 14.4 | 3.7 |
| | 713 | 14.0 | 1.0 |
| | 734 | 13.6 | 1.7 |
| | 765 | 13.1 | 5.1 |
| | 805 | 12.4 | 2.9 |
| | 917 | 10.9 | 6.2 |
| | 927 | 10.8 | 6.1 |
| | 952 | 10.5 | 6.3 |
| | 1009 | 9.9 | 137.9 |
| | 1070 | 9.3 | 24.5 |
| | 1088 | 9.2 | 3.6 |
| | 1128 | 8.9 | 5.3 |
| | 1169 | 8.6 | 27.2 |
| | 1195 | 8.4 | 17.1 |
| | 1224 | 8.2 | 12.5 |
| | 1240 | 8.1 | 11.5 |
| | 1255 | 8.0 | 12.0 |
| | 1262 | 7.9 | 10.8 |
| | 1270 | 7.9 | 11.4 |
| | 1278 | 7.8 | 32.4 |
| | 1288 | 7.8 | 19.0 |
| | 1305 | 7.7 | 24.3 |
| | 1328 | 7.5 | 15.6 |
| | 1369 | 7.3 | 24.2 |
| | 1401 | 7.1 | 14.1 |
| | 1419 | 7.0 | 16.6 |
| | 1433 | 7.0 | 16.9 |
| | 1473 | 6.8 | 20.6 |
| | 1495 | 6.7 | 12.2 |
| | 1521 | 6.6 | 23.1 |
| | 1548 | 6.5 | 10.7 |
| $C_{70}H^+$ | 461 | 21.7 | 6.7 |
| | 538 | 18.6 | 7.8 |
| | 572 | 17.5 | 5.1 |
| | 637 | 15.7 | 4.0 |
| | 660 | 15.2 | 4.5 |
| | 680 | 14.7 | 1.7 |
| | 697 | 14.3 | 1.0 |
| | 724 | 13.8 | 1.3 |
| | 731 | 13.7 | 1.3 |
| | 798 | 12.5 | 1.6 |
| | 891 | 11.2 | 1.2 |
| | 928 | 10.8 | 3.0 |
| | 1012 | 9.9 | 3.5 |
| | 1040 | 9.6 | 3.5 |
| | 1057 | 9.5 | 1.3 |
| | 1069 | 9.4 | 5.4 |
| | 1086 | 9.2 | 5.0 |
| | 1103 | 9.1 | 4.9 |
| | 1135 | 8.8 | 5.5 |
| | 1163 | 8.6 | 5.3 |
| | 1186 | 8.4 | 6.4 |
| | 1216 | 8.2 | 4.0 |
| | 1237 | 8.1 | 6.4 |
| | 1250 | 8.0 | 5.0 |
| | 1276 | 7.8 | 7.0 |
| | 1301 | 7.7 | 4.5 |
| | 1340 | 7.5 | 8.8 |
| | 1367 | 7.3 | 4.3 |
| | 1380 | 7.2 | 10.2 |
| | 1414 | 7.1 | 14.6 |
| | 1432 | 7.0 | 24.0 |
| | 1453 | 6.9 | 15.1 |
| | 1469 | 6.8 | 7.1 |
| | 1491 | 6.7 | 9.8 |
| | 1529 | 6.5 | 20.4 |
| $C_{60}V^+$ | 417 | 24.0 | 1.8 |
| | 445 | 22.5 | 5.3 |
| | 511 | 19.6 | 5.2 |
| | 522 | 19.2 | 12.0 |
| | 582 | 17.2 | 11.3 |
| | 605 | 16.5 | 1.3 |
| | 697 | 14.3 | 1.1 |
| | 719 | 13.9 | 1.3 |
| | 952 | 10.5 | 1.1 |
| | 1081 | 9.3 | 2.1 |
| | 1168 | 8.6 | 4.7 |
| | 1177 | 8.5 | 7.8 |
| | 1190 | 8.4 | 4.5 |
| | 1235 | 8.1 | 1.7 |
| | 1283 | 7.8 | 2.3 |
| | 1305 | 7.7 | 1.0 |
| | 1323 | 7.6 | 1.4 |
| | 1378 | 7.3 | 6.3 |
| | 1406 | 7.1 | 7.1 |
| | 1411 | 7.1 | 7.4 |
| | 1426 | 7.0 | 16.7 |
| | 1469 | 6.8 | 6.7 |
| | 1500 | 6.7 | 6.3 |
| | 1517 | 6.6 | 17.2 |
| | 1529 | 6.5 | 7.6 |
| | 1548 | 6.5 | 1.5 |

In the JWST era, the high-sensitivity, high-resolution infrared spectral data would provide an unprecedented opportunity to identify specific species and to determine their abundances in space, or put upper limits on their abundances in case of nondetection. To this end, Table 4 summarizes the scaled harmonic frequencies and corresponding transition intensities of the fullerene derivatives. These data would be useful for exploring their cosmic relevance in comparison with astronomical data, in particular the upcoming higher quality JWST data. Nevertheless, it should be noted that conclusive identification of a specific species in space via those frequencies is still challenging, as pointed out previously (García-Hernández et al. 2010; Kwok, 2022; Hou, et al. 2023). For example, the features of fullerene derivatives at 7.7, 11.3, and 16.4 µm as shown in Table 4 could potentially coincide with the UIE bands often attributed to PAHs (Kwok 2022) by taking into account of the frequency uncertainty. In addition, some features of the fullerene derivatives might even be hidden within the broad 6–9 and 11–13 µm emission plateaus (García-Hernández et al. 2010; García-Hernández et al. 2011; Kwok 2022; Hou, et al. 2023). However, it might be advisible to realize that one should anayze the whole set of vibrational modes and their frequencies of a particular specises to understand its cosmic relevance and importance, instead of relying on a specific band feature (Hou, et al. 2023). With the unprecedented resolution and sensitivity of JWST in the 4.9 to 28.3 µm spectral range, distinguishing these species would become possible. It would be expected that such in-depth knowledge about the distinctive infrared signatures of these carbon species would undoubtedly contribute to our understanding of the organic inventory and carbon evolution in the universe.

## 6. CONCLUSIONS

We have conducted a systematic evaluation on the performance of common DFT functionals in predicting the infrared spectra of fullerenes and their derivatives. We have derived frequency-range-specific scaling factors for the functionals of BP86, BPW91, TPSSh, B3LYP, PBE0, M06, M06-2X, and ωB97XD, all with 6-31G** basis set. Overall, BP86 and BPW91 are recommended for simulating the infrared spectra of $C_{60}$ and its derivatives regarding both their accuracy and computational efficiency. It is found that the coupling modes between the fullerene cage and the binding atom, in general are challenging to be accurately calculated. We also validated the reliability of the fitted factors by comparing the scaled theoretical and experimental infrared spectra of $[C_{60}\text{-Metal}]^+$ complexes, highlighting the usefulness of the current work and warranting the reliable prediction of potential fullerene-related species important in space. It is expected that the current result would greatly facilitate the analysis and interpretation of the upcoming JWST data particularly in the mid-infrared region covered by the Mid-Infrared Instrument (MIRI) on board JWST.

### Author Contributions

G.-L.H. conceived and designed the work, J.X. conducted the work, J.X., G.-L.H., and A.L. wrote the manuscript with comments from all authors, and G.-L.H. obtained the funding for this research. All authors have given approval to the final version of the manuscript.

### Notes

The authors declare no competing financial interests.

### ACKNOWLEDGEMENTS

This work was supported by National Natural Science Foundation of China (92261101) and the Innovation Capability Support Program of Shaanxi Province (2023-CX-TD-49). The authors acknowledge the support from the “Young Talent Support Plan” of Xi’an Jiaotong University, the "Excellent Overseas Young Scientist Program", "Selective Funding Project for Overseas Student Scientific and Technological Activities" of Shaanxi province, China, as well as the Fundamental Research Funds for Central Universities, China. A.L. is supported in part by NASA grants (S0NSSC19K0572 and S0NSSC19K0701). Part of the computational resources and services used in this work were provided by the VSC (Flemish Supercomputer Center), funded by the Research Foundation-Flanders (FWO) and the Flemish Government-department EWI.

**DATA AVAILABILITY**

The data underlying this article is available in the article. Any inquiry about further details can be addressed to G.-L.H.

## APPENDIX

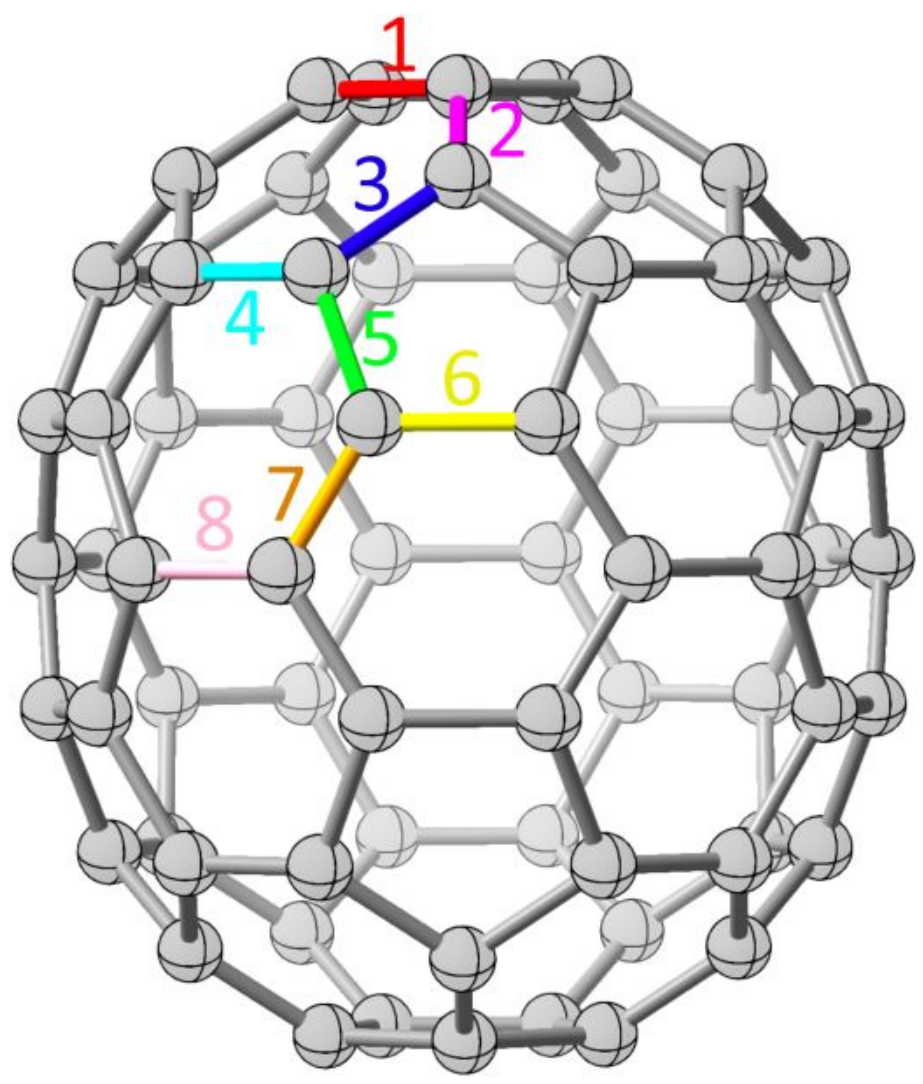

**Figure A1.** Notation of the C70 characteristic bonds. Geometry of $C_{70}$, with 8 characteristic bonds in different colors.

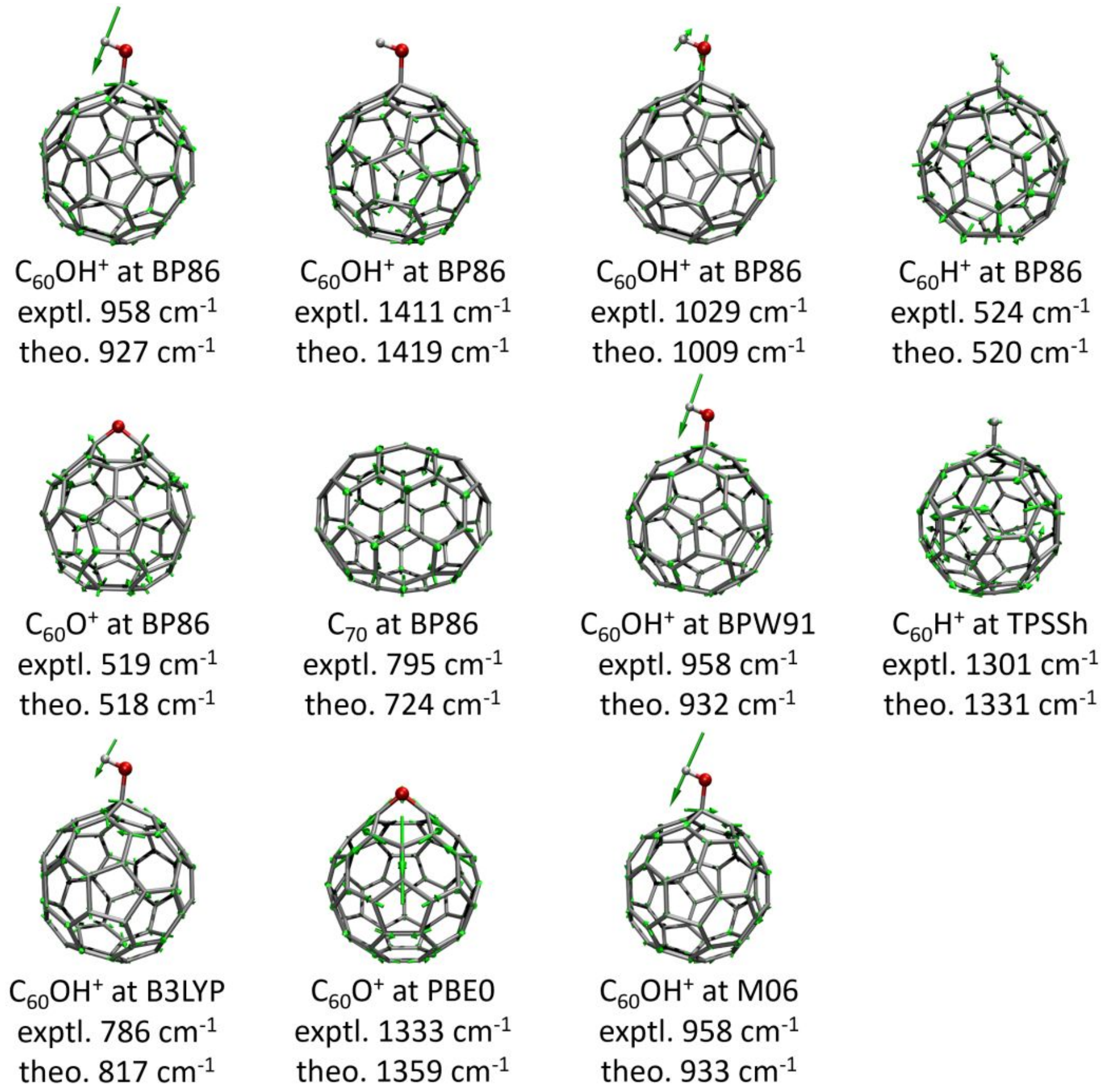

**Figure A2.** Selected vibrational vectors of modes at specific functional.

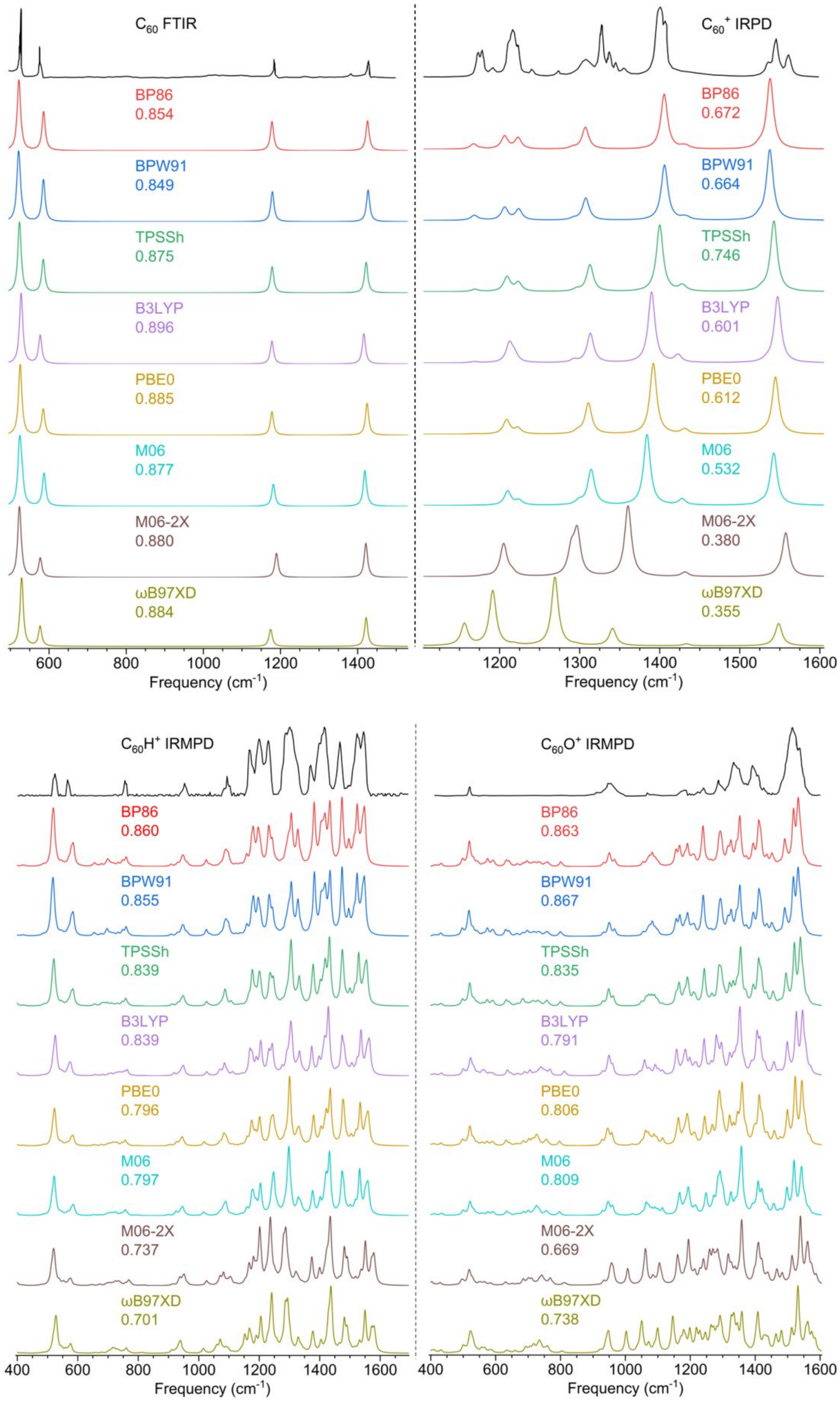

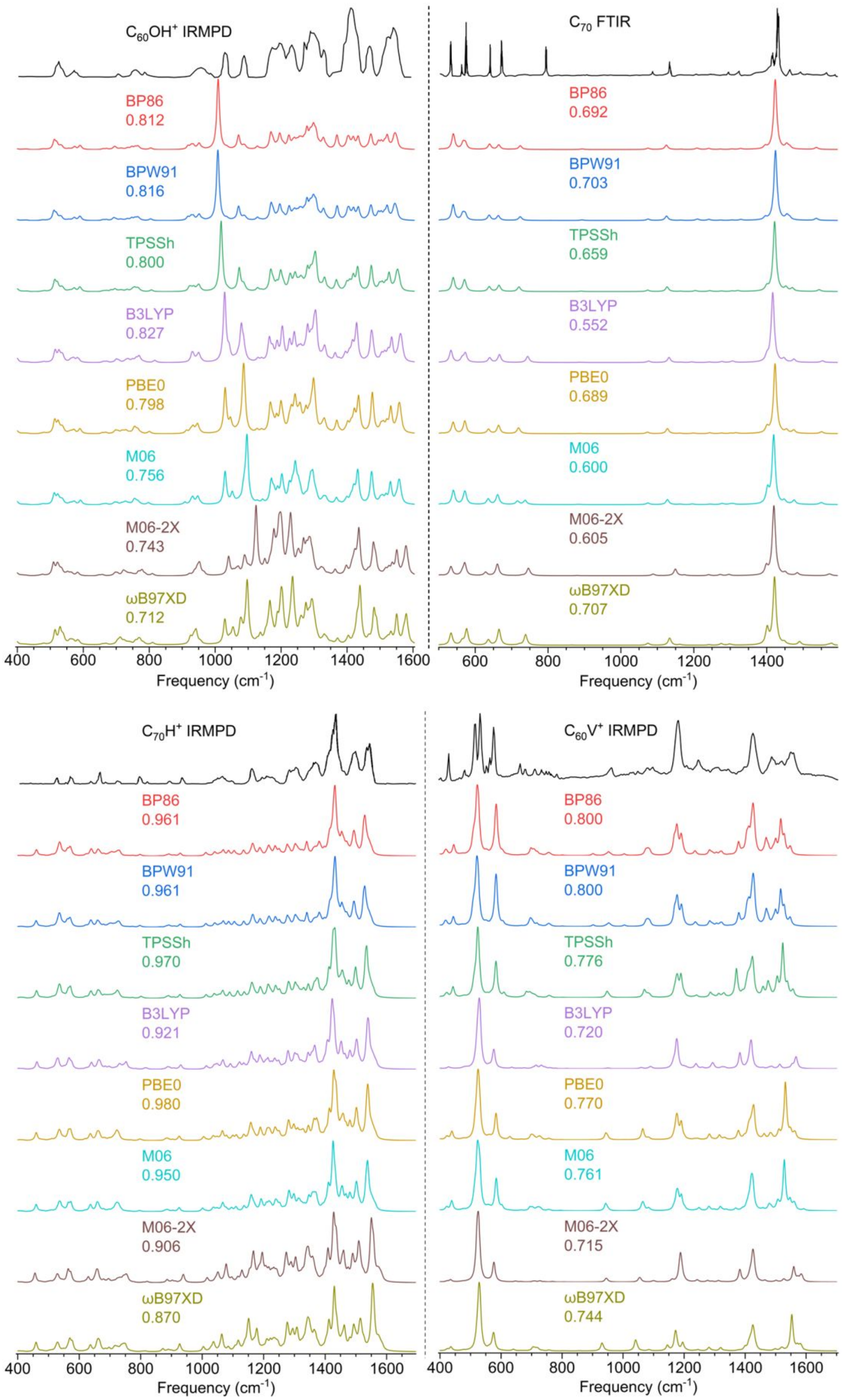

**Figure A3**. Performance of DFT functionals in predicting the infrared spectra. Comparison of scaled spectra with experimental measurements of $C_{60}$, $C_{60}^{+}$, $C_{60}H^{+}$, $C_{60}O^{+}$, $C_{60}OH^{+}$, $C_{70}$, $C_{70}H^{+}$, $C_{60}V^{+}$. The spectra are convolved using Lorentzian line shapes of 10 $cm^{-1}$ full width at half maximum, and calculated Cosine similarity score for each functional is indicated.

**Table A1.** Fitting goodness of specific DFT functional. RMSE/median error values fitted using frequency-range-specific and global fitting strategies. The segmentation point of 50 $cm^{-1}$ in the whole spectral range is used to determine the best fitting results for a specific functional.

| Seg. point | BP86 | BPW91 | TPSSh | B3LYP | PBE0 | M06 | M06-2X | ωB97XD |
|---|---|---|---|---|---|---|---|---|
| 550 | 11.6/6.4 | 11.9/6.8 | 11.6/7.2 | 11.3/7.5 | 11.5/7.9 | 12.7/8.2 | 18.7/12.2 | 17.2/9.7 |
| 600 | 11.3/5.7 | 11.8/6.3 | 11.2/6.5 | 11.2/6.9 | 11.0/6.9 | 12.2/7.7 | 18.6/12.3 | 17.0/9.1 |
| 650 | 10.8/5.3 | 11.3/5.8 | 10.9/6.4 | 11.2/6.9 | 11.0/6.4 | 11.7/7.9 | 18.6/12.3 | 17.0/9.1 |
| 700 | 10.2/5.3 | 10.7/5.5 | 10.1/6.3 | 11.1/5.6 | 10.0/6.4 | 11.0/7.5 | 18.4/11.0 | 16.8/8.5 |
| 750 | 10.4/5.6 | 10.9/5.7 | 10.3/6.3 | 11.0/5.2 | 9.6/6.3 | 10.6/7.0 | 18.3/10.8 | 16.8/8.5 |
| 800 | 10.8/6.7 | 11.3/6.9 | 10.6/7.2 | 11.3/6.6 | 10.2/7.6 | 11.3/8.2 | 17.9/11.6 | 17.1/8.7 |
| 850 | 11.4/7.7 | 11.7/8.5 | 11.2/8.3 | 11.4/7.2 | 10.8/8.4 | 11.8/8.9 | 17.8/10.7 | 17.2/8.9 |
| 900 | 11.4/7.9 | 11.7/8.5 | 11.2/8.3 | 11.4/7.2 | 10.8/8.4 | 11.8/8.9 | 18.0/10.6 | 17.3/10.3 |
| 950 | 11.2/8.1 | 11.5/8.5 | 11.3/8.3 | 11.4/7.2 | 10.8/8.8 | 11.9/9.2 | 17.9/10.4 | 17.2/10.4 |
| 1000 | 11.3/8.1 | 11.7/8.2 | 11.1/8.3 | 11.4/7.2 | 10.7/8.4 | 11.6/9.2 | 17.8/9.6 | 17.1/9.2 |
| 1050 | 11.0/7.4 | 11.4/7.8 | 11.1/7.8 | 11.4/7.2 | 10.7/8.4 | 11.6/9.2 | 17.8/9.6 | 17.1/9.2 |
| 1100 | 11.4/7.6 | 11.8/7.6 | 11.1/7.7 | 11.3/6.9 | 10.6/8.2 | 11.6/9.4 | 17.5/9.9 | 16.9/9.5 |
| 1150 | 11.4/7.4 | 11.8/7.2 | 11.1/7.6 | 11.3/6.6 | 10.8/8.3 | 11.8/9.8 | 16.9/9.6 | 16.5/8.6 |
| 1200 | 11.8/7.2 | 12.0/6.2 | 11.3/7.6 | 11.3/6.7 | 10.8/8.3 | 12.4/9.0 | 17.4/9.4 | 16.7/9.5 |
| 1250 | 12.0/6.4 | 12.3/6.8 | 11.6/7.5 | 10.9/7.3 | 11.1/8.0 | 12.7/8.7 | 16.4/8.3 | 16.6/8.9 |
| 1300 | 11.8/6.1 | 12.3/6.6 | 11.9/7.5 | 11.3/7.0 | 11.4/8.0 | 13.0/8.5 | 16.5/10.3 | 16.4/9.5 |
| 1350 | 12.1/7.1 | 12.4/7.8 | 11.8/7.7 | 10.6/6.3 | 11.6/8.2 | 13.1/9.0 | 16.6/10.3 | 15.7/10.6 |
| 1400 | 12.2/6.7 | 12.5/7.5 | 12.1/8.2 | 11.3/7.0 | 11.5/7.6 | 13.1/9.6 | 15.6/10.2 | 16.1/9.5 |
| 1450 | 12.2/8.5 | 12.5/7.9 | 11.7/7.4 | 10.9/6.4 | 11.7/7.6 | 13.2/10.0 | 15.0/10.3 | 12.2/9.9 |
| 1500 | 12.1/7.5 | 12.4/7.5 | 12.1/8.3 | 11.2/6.5 | 11.6/7.0 | 12.8/8.7 | 15.4/10.0 | 15.1/9.6 |
| 1550 | 12.1/7.8 | 12.4/7.2 | 12.1/8.9 | 12.3/7.3 | 11.9/7.2 | 13.2/9.8 | 15.8/9.2 | 15.2/8.9 |
| Global | 12.3/7.0 | 12.5/7.4 | 12.1/8.0 | 11.4/7.3 | 12.0/7.9 | 13.3/9.4 | 18.8/12.5 | 17.3/9.6 |